\documentclass[a4paper,11pt]{article}
\usepackage[utf8]{inputenc}
\usepackage{amsfonts}
\usepackage{amsmath}
\usepackage{amssymb}
\usepackage{indentfirst}
\usepackage{graphicx}
\usepackage[colorlinks]{hyperref}
\usepackage{cite}
\usepackage{color}
\usepackage{xcolor}
\usepackage{ulem,lipsum}
\usepackage{soul} 
\usepackage{breqn}

\usepackage{setspace}%

\usepackage[symbol]{footmisc}

\renewcommand{\thefootnote}{\fnsymbol{footnote}}
\numberwithin{equation}{section}
\allowdisplaybreaks
\makeatletter

\begin{document}

\begin{titlepage}
\rightline{  }
\vspace{1cm}

\baselineskip=24pt

\begin{center}
\textbf{\LARGE Four dimensional topological supergravities from transgression field theory}
\par\end{center}{\LARGE \par}

\begin{center}
	\vspace{0.5cm}
	Patrick Concha,$^{1,2,}\footnote[1]{patrick.concha@ucsc.cl}$
		Fernando Izaurieta,$^{3,}\footnote[2]{fernando.izaurieta@uss.cl}$
			Evelyn Rodr\'{i}guez,$^{1,2,}\footnote[3]{erodriguez@ucsc.cl}$   and
				Sebasti\'{a}n Salgado$^{4,}\footnote[4]{sebasalg@gmail.com}$
	\small
	\\[4mm]

	$^{1}$\textit{Departamento de Matem\'{a}tica y F\'{i}sica Aplicadas, Universidad Cat\'{o}lica de la Sant\'{i}sima Concepci\'{o}n,\\
	Alonso de Ribera 2850, Concepci\'{o}n, Chile}
		\\[4mm]

  $^{2}$\textit{Grupo de Investigación en Física Teórica, GIFT,\\
  Concepción, Chile.}
		\\[4mm]

		$^{3}$\textit{Facultad de Ingenier\'{i}a, Arquitectura y Diseño, Universidad San Sebasti\'{a}n,\\
Lientur 1457, Concepción 4080871, Chile}
	\\[4mm]
	
	$^{4}$\textit{Sede Esmeralda, Universidad de Tarapac\'{a},\\ Av. Luis Emilio Recabarren 2477, Iquique, Chile}

\vspace{0.5cm}
 
	\par\end{center}
\vskip 20pt
\begin{abstract}
\noindent In this work, we propose a four-dimensional gauged Wess-Zumino-Witten model,
obtained as a dimensional reduction from a transgression field theory
invariant under the $\mathcal{N}=1$ Poincar\'{e} supergroup. For this
purpose, we consider that the two gauge connections on which the
transgression action principle depends are given by linear and non-linear
realizations of the gauge group respectively. The field content of the
resulting four-dimensional theory is given by the gauge fields of the linear
connection, in addition to a set of scalar and spinor multiplets in the
same representation of the gauge supergroup, which in turn, correspond to
the coordinates of the coset space between the gauge group and the
five-dimensional Lorentz group. We then decompose the action in terms of
four-dimensional quantities and derive the corresponding equations of
motion. We extend our analysis to the non- and ultra- relativistic regime.

	\end{abstract}
\end{titlepage}\newpage {} 
\noindent\rule{162mm}{0.4pt}
\tableofcontents
\noindent\rule{162mm}{0.4pt}
\renewcommand*{\thefootnote}{\arabic{footnote}}

\section{Introduction}

In the last decades, several gravitational theories have been introduced as
alternatives to General Relativity. The most general theory that can be
formulated in arbitrary dimensions and fulfill the fundamental requirements
of invariance under diffeomorphisms, lead to equations of motion of second
degree in the metric tensor and preserve the conservation law of the
energy-momentum tensor is known as the Lovelock theory \cite%
{Lanczos:1938sf,Lovelock:1971yv}. The Lovelock Lagrangian density is a sum
over all the possible combinations between the Lorentz curvature depending
on the spin connection, and the vielbein that codifies the metricity. In three and four dimensions, the Lovelock theory
reproduces General Relativity with positive, negative, or zero cosmological
constant depending on the values of its arbitrary constants \cite%
{Zumino:1985dp,Teitelboim:1987zz,Mardones:1990qc,Crisostomo:2000bb}. A case
of special interest is obtained when the constants of the sum in the
Lovelock Lagrangian are fixed such that the theory presents the maximum
number of degrees of freedom \cite{Troncoso:1999pk}. In that case, the
three-dimensional Lovelock Lagrangian becomes proportional to the
Chern--Simons (CS) three-form of the AdS group \cite%
{Assimos:2019yln,Banados:1996hi}, which has a topological origin and is, up
to boundary terms, invariant under the gauge transformations induced by the
AdS group.  Such a feature allows us to formulate General Relativity as
a topological gauge theory in three dimensions. However, the same does not
occur in four dimensions, since CS forms exist only in odd
dimensions. Moreover, the CS form of the AdS group does not
lead to General Relativity in dimensions higher than three in any regime.
With the purpose of formulating even-dimensional gravity theories,
especially in four dimensions, A. H. Chamseddine proposed a topological
gravity depending on the same variables than the Lovelock theory, in
addition to a scalar multiplet in the same representation of the gauge group 
\cite{Chamseddine:1990gk,Chamseddine:1989nu,Chamseddine:1989yz}.

From a mathematical point of view, CS forms appear in the study
of topological invariant densities that only exist in even dimension and, as
a consequence of the Poincar\'{e} lemma, allow the existence odd-dimensional
secondary forms that inherit their gauge invariance properties. These forms
known as transgression forms, become CS forms as locally defined
particular cases \cite%
{Mora:2003wy,Mora:2006ka,Izaurieta:2005vp,Borowiec:2005ky,Izaurieta:2006wv,Izaurieta:2006aj}%
. Thus, both transgression and CS forms are often used as
Lagrangian densities for odd-dimensional gauge theories. In contrast to
CS theories, transgression field theories depend on two gauge
connections and their Lagrangian densities are globally defined differential
forms that, as well as the topological densities in which they originate,
are fully gauge invariant. In addition to be useful in the construction of
Lagrangian densities, transgression forms can naturally induce a dimensional
reduction that allows to formulate even-dimensional gauge invariant theories
without the need of breaking the gauge covariance. These are known gauged
Wess--Zumino--Witten (gWZW) models \cite{Salgado:2014iha,Salgado:2013pva}
and, as well as the even-dimensional topological gravity proposed by
Chamseddine, include a scalar multiplet in the gauge group representation as
part of the fundamental field content. Indeed, it was shown in \cite{Salgado:2014iha,Inostroza:2014vua,Salgado:2013pva} that
Chamsedine's topological gravities can be obtained as gWZW models of the
Poincar\'{e} group in arbitrary even dimensions. Furthermore, in ref.\cite%
{Salgado:2013pva}, these results were generalized to the Maxwell algebra and
the Poincar\'{e} superalgebra in three dimensions, obtaining fully gauge
invariant ($1+1$)-dimensional theories for gravity and supergravity
respectively. Moreover, in refs. \cite{Salgado:2014iha,Salgado:2013pva}, it
was studied the relation between four-dimensional gWZW models and general
relativity.

The existence of abstract even-dimensional gWZW models motivates us to study
a four-dimensional gauge invariant supergravity theory that emerges from
considering a supersymmetric extension of the five-dimensional Poincar\'{e}
supergroup as gauge group. Moreover, due to the well-known relation between
the Poincar\'{e} group and the Galilei and Carroll groups \cite%
{Duval:2014uoa}, we aim the purpose of studying the non- and ultra-
relativistic regimes of the resulting theory and thus to obtain the
corresponding superalgebras and gWZW action principles.

This paper is organized as follows: In section \ref{sec2}, we consider a brief
introduction to the SW-GN formalism and the gWZW model. In section \ref{sec3}, we
study the non-linear realization of the five-dimensional Poincar\'{e}
superalgebra. In section \ref{sec4} and \ref{sec5}, we derive the four-dimensional gWZW action
invariant under the aforementioned Poincar\'{e} supergroup, and study the
dynamics of the resulting theory. In section \ref{sec6}, we consider the
non-relativistic limit of the Poincar\'{e} superalgebra and derive the
corresponding gWZW non-relativistic action principle. In section \ref{sec7} we
perform the same analysis for the ultra-relativistic limit of the theory.
Section \ref{ccl} contains our final conclusions.

\section{Non-linear realizations and gWZW actions}\label{sec2}

In this section we briefly review the Stelle-West-Grignani-Nardelli (SW-GN)
formalism and the gauged Wess-Zumino-Witten (gWZW) models.

\subsection{The SW-GN formalism}

The Stelle-West-Grignani-Nardelli formalism makes use of non-linear
realizations of Lie groups in the construction of gauge invariant action
principles \cite{Stelle:1979aj,Grignani:1991nj}. Thus, the gauge symmetry of
a physical theory can be extended from a stability group, to a
higher-dimensional group that contains it as a subgroup. Let us consider a
Lie group $G$ with Lie algebra $\mathcal{G}$, and a subgroup $H\subset G$
with Lie algebra $\mathcal{H}$ as stability subgroup. We denote as $\left\{
V_{i}\right\} _{i=1}^{\dim H}$ to the basis of $\mathcal{H}$ and as $\left\{
T_{l}\right\} _{l=1}^{\dim G-\dim H}$ the to the set of generators of the
remaining subspace. We assume that the basis can be chosen such that the
generators $T_{l}$ form a representation of the stability subgroup.
Therefore, the Lie products between the vectors of the introduced basis
satisfy $\left[ V,T\right] \backsim T$, i.e. these products are linear
combinations of $T_{l}$. An arbitrary group element $g$ can be decomposed in
terms the generators of the subgroup and the remaining subspace as%
\begin{equation}
g=e^{\xi ^{l}T_{l}}h\,,  \label{1}
\end{equation}%
where $h\in H$ is a group element defined by the action of the group on the
zero-forms $\xi ^{l}$ which, in turn, play the role of coordinates that
parametrize the coset space $G/H$. From eq. (\ref{1}), it follows that the
action of an arbitrary element $g_{0}\in G$ on $e^{\xi ^{l}T_{l}}$ can be
also split as 
\begin{equation}
g_{0}e^{\xi ^{l}T_{l}}=e^{\xi ^{\prime l}T_{l}}h_{1}\,.  \label{law}
\end{equation}%
Eq. (\ref{law}) allows to obtain the non-linear functions $\xi ^{\prime
}=\xi ^{\prime }\left( g,\xi \right) $ and $h_{1}=h_{1}\left( g,\xi \right) $%
. By considering that the transformation law $\xi \rightarrow \xi ^{\prime }$
is described by the variation $\delta $, and by choosing the group element $%
g_{0}$ such that $\left( g_{0}-1\right) $ is infinitesimal, eq. \eqref{law}
leads to \cite{Coleman:1969sm,Callan:1969sn,Salam:1969rq} 
\begin{equation}
e^{-\xi ^{l}T_{l}}\left( g_{0}-1\right) e^{\xi ^{l}T_{l}}-e^{-\xi
^{l}T_{l}}\delta e^{\xi ^{l}T_{l}}=h_{1}-1\,.  \label{5}
\end{equation}%
Since $g-1$ is infinitesimal, $h_{1}-1$ is a vector of $H$.

Let us now consider the case in which $g_{0}=h_{0}$ belongs to the stability
subgroup. In this case, eq. \eqref{law} becomes 
\begin{equation}
e^{\xi ^{\prime l}T_{l}}=\left( h_{0}e^{\xi ^{l}T_{l}}h_{0}^{-1}\right)
h_{0}h_{1}^{-1}\,,  \label{6}
\end{equation}%
and since the Lie product $\left[ V_{i},T_{l}\right] $ is proportional to $%
T_{l}$, one gets $h_{0}=h_{1}$ and the transformation law becomes linear: 
\begin{equation}
e^{\xi ^{\prime l}T_{l}}=h_{0}e^{\xi ^{l}T_{l}}h_{0}^{-1}\,.
\end{equation}%
On the other hand, if we consider $g_{0}=e^{\xi _{0}^{l}T_{l}}$, eq. \eqref{law} becomes 
\begin{equation}
e^{\xi ^{\prime l}T_{l}}=e^{\xi _{0}^{l}T_{l}}e^{\xi ^{l}T_{l}}h^{-1}\,,
\end{equation}%
which is a non-linear transformation law for $\xi $.

Let us now consider a one-form gauge connection $A$ taking values on $%
\mathcal{G}$ and an action principle $S=S\left[ A\right] $ with gauge
invariance under the transformations of the stability subgroup $\mathcal{H}$
but not under those along the generators of the coset space. Under the
action of an arbitrary group element $g$, the gauge connection transforms as 
\cite{Stelle:1979aj,Grignani:1991nj,Salgado:2001bn,Salgado:2003rf} 
\begin{equation}
A\longrightarrow A^{\prime }=g^{-1}dg+g^{-1}Ag\,.  \label{11}
\end{equation}%
We split $\mu $ into its contributions belonging in $h$ and the coset space
as $A=a+\rho $, with $a=a^{l}T_{l}$ and $\rho =\rho _{i}V_{i}$. Moreover, we
introduce a group element $z=\exp \left( \xi ^{l}A_{l}\right) $ and define
the non-linear gauge connection%
\begin{equation}
A^{z}=z^{-1}\text{d}z+z^{-1}Az\,.  \label{12}
\end{equation}%
The functional form of $A^{z}$ is given by a large gauge transformation of $%
A $ that non-linearly depends on the zero-forms $\xi ^{l}$ and their
derivatives. However, in the SW-GN formalism $A^{z}$ is interpreted as the
fundamental field of a gauge theory and therefore, both $A$ and $\xi $ will
change under the action of the gauge group. As before, we split the
contributions to the non-linear connection as $A^{z}=v+p$ with%
\begin{eqnarray}
p &=&p^{l}\left( \xi ,\text{d}\xi \right) T_{l}\,, \notag\\
v &=&v^{i}\left( \xi ,\text{d}\xi \right) V_{i}\,.
\end{eqnarray}%
It is possible to prove that, under the transformation $\delta $ generated
by the action of the group, the transformation laws for $p$ and $v$ are
given by%
\begin{align}
p& \longrightarrow p^{\prime }=h_{1}^{-1}ph_{1}\,,  \notag \\
v& \longrightarrow v^{\prime }=h_{1}^{-1}vh_{1}+h_{1}^{-1}\text{d}h_{1}\,,
\label{v}
\end{align}%
i.e., when acting with a group element belonging to the coset space, the
non-linear one-forms $p$ and $v$ transform as a tensor and as a connection
respectively. These transformations are linear but the group element is now
a function of the parameters $h_{1}=h_{1}\left( \xi _{0},\xi \right) $. From
the transformations laws in eq. \eqref{v}, it follows that
the non-linear gauge connection transforms in the same way that under the
action of the stability subgroup and the coset space. Therefore, an action
principle defined as a functional of $A$ whose gauge symmetry is described
by the stability subgroup, becomes invariant under the entire group $G$ when 
$A$ is replaced by $A^{z}$. The original non-invariance of the action
principle is thus compensated by the transformation law of the gauge
parameters $\xi $.

\subsection{gWZW models}

Let us consider two independent gauge connections $A_{1}$ and $A_{2}$
evaluated in the same gauge algebra. The transgression $\left( 2n+1\right) $%
-form corresponding to both gauge connections is defined as%
\begin{equation}
Q_{A_{2}\leftarrow A_{1}}^{\left( 2n+1\right) }=\left( n+1\right)
\int_{0}^{1}\left\langle \left( A_{2}-A_{1}\right) F_{t}^{n}\right\rangle\,,
\end{equation}%
where $\left\langle \text{ }\right\rangle $ denotes the symmetrized trace
along the generators of the Lie algebra, and $F_{t}$ is the gauge curvature
associated to the homotopic gauge connection $A_{t}=A_{1}+t\left(
A_{2}-A_{1}\right) $. Transgression forms are globally defined and fully
invariant under the transformations of the gauge group. CS forms
emerge as particular cases of transgression forms, by locally setting one of
the gauge connections as vanishing. Thus, the CS form
corresponding to a gauge connection $A$ is locally defined as%
\begin{equation}
Q_{A\leftarrow 0}^{\left( 2n+1\right) }=\left( n+1\right)
\int_{0}^{1}\left\langle AF_{t}^{n}\right\rangle\,,
\end{equation}%
where the homotopic gauge connection takes form $A_{t}=tA$. Furthermore, by
applying the Cartan homotopy formula, it is possible to prove that a general
transgression form can be written in terms of two CS forms and a
total derivative, as follows \cite{Manes:1985df,Izaurieta:2005vp,Mora:2006ka}%
\begin{equation}
Q_{A_{2}\leftarrow A_{1}}^{\left( 2n+1\right) }=Q_{A_{2}\leftarrow
A_{0}}^{\left( 2n+1\right) }-Q_{A_{1}\leftarrow A_{0}}^{\left( 2n+1\right)
}-Q_{A_{2}\leftarrow A_{1}\leftarrow A_{0}}^{\left( 2n\right) }\,.  \label{key}
\end{equation}%
The $2n$-form inside the exterior derivative is explicitly given as the
following integral: 
\begin{equation}
Q_{A_{2}\leftarrow A_{1}\leftarrow A_{0}}^{\left( 2n\right) }=n\left(
n+1\right) \int_{0}^{1}\text{d}t\int_{0}^{t}\text{d}s\left\langle \left(
A_{2}-A_{1}\right) \left( A_{1}-A_{0}\right) F_{st}^{n-1}\right\rangle \,.
\label{bound}
\end{equation}%
where $F_{st}$ is the gauge curvature associated to the homotopic gauge field%
\begin{equation}
A_{st}=A_{0}+t\left( A_{1}-A_{0}\right) +s\left( A_{2}-A_{1}\right) \,,
\end{equation}%
which depends on two parameters $t$ and $s$ taking values between $0$ and $1$%
. For details on the use of the Cartan homotopy formula and the homotopy
operator in this context, see refs. \cite%
{Nakahara:2003nw,Izaurieta:2005vp,Mora:2006ka,Izaurieta:2006wv}.

Let us now consider two gauge connections $A$ and $A^{z}$, related by the
gauge transformation $A^{z}=z^{1}\left( \text{d}+A\right) z$, where $z=\exp
\left( \xi \right) $ is an element of the gauge group and $\xi $ a zero-form
mutiplet in the same representation of the Lie algebra. From eq. (\ref{rules}%
), it follows that the transgression form associated to $A$ and $A^{z}$ can
be written in terms of the difference between their corresponding
CS forms. Let us now introduce a homotopic gauge field $A_{t}=tA$
which takes values between $0$ and $A$, as the parameter $t$ takes values
between $0$ and $1$. The transformed connection, obtained from $A_{t}$,
denoted by $\left( A_{t}\right) ^{z}$, and its corresponding gauge curvature 
$\left( F_{t}\right) ^{z}$ are given by\footnote{%
Note that in general $\left( A_{t}\right) ^{z}\neq \left( A^{z}\right) _{t}$}%
\begin{eqnarray}
\left( A_{t}\right) ^{z} &=&z^{-1}tAz+z^{-1}\text{d}z\,, \notag\\
\left( F_{t}\right) ^{z} &=&z^{-1}F_{t}z=z^{-1}\left( tF+\left(
t^{2}-t\right) A^{2}\right) z\,.
\end{eqnarray}%
These homotopic quantities verify%
\begin{eqnarray}
\left( A_{0}\right) ^{z} &=&z^{-1}\nu z=z^{-1}\text{d}z,\text{ \ \ \ \ }%
\left( A_{1}\right) ^{z}=A^{z}\,, \notag\\
\left( F_{0}\right) ^{z} &=&0,\text{ \ \ \ \ }\left( F_{1}\right) ^{z}=F^{z}\,.
\end{eqnarray}%
By applying the Cartan homotopy formula once again, it is possible to prove
that the CS forms corresponding to the pure gauge connection $%
z^{-1}$d$z$ and the transformed gauge connections $A^{z}$ are related by the
following equation%
\begin{equation}
Q_{A^{z}\leftarrow 0}^{\left( 2n+1\right) }-Q_{z^{-1}dz\leftarrow 0}^{\left(
2n+1\right) }=\left( k_{01}\text{d}+\text{d}k_{01}\right) Q_{\left(
A_{t}\right) ^{z}\leftarrow 0}^{\left( 2n+1\right) }\,,  \label{hom}
\end{equation}%
with $k_{01}=\int_{0}^{1}\ell _{t}$, where $\ell _{t}$ is the homotopy
operator defined by the following action on $A_{t}$ and $F_{t}$:%
\begin{equation}
\ell _{t}A_{t}=0,\text{ \ \ \ \ \ }\ell _{t}F_{t}=\text{d}t\frac{\partial }{%
\partial t}A_{t}\,.  \label{rules}
\end{equation}%
By directly applying eq. (\ref{rules}), one finds that the first term in the
right side of (\ref{hom}) is given by%
\begin{equation}
k_{01}\text{d}Q_{\left( A_{t}\right) ^{z}\leftarrow 0}^{\left( 2n+1\right)
}=Q_{A\leftarrow 0}^{\left( 2n+1\right) }\,,
\end{equation}%
so that, eq. (\ref{hom}) allows to write the difference between two
CS forms related by means of a gauge connection as%
\begin{equation}
Q_{A^{z}\leftarrow 0}^{\left( 2n+1\right) }-Q_{A\leftarrow 0}^{\left(
2n+1\right) }=Q_{z^{-1}\text{d}z\leftarrow 0}^{\left( 2n+1\right) }+\text{d}%
\alpha _{2n}\left( A,z\right)\,,  \label{dif}
\end{equation}%
where we introduce%
\begin{equation}
\alpha _{2n}\left( A,z\right) =k_{01}Q_{\left( A_{t}\right) ^{z}\leftarrow
0}^{\left( 2n+1\right) }\,.
\end{equation}%
Notice that the first term in the r.h.s. of eq. (\ref{dif}) is the
CS form corresponding to the pure gauge connection $z^{-1}$d$z$,
which is explicitly given by 
\begin{equation}
Q_{z^{-1}\text{d}z\leftarrow 0}^{\left( 2n+1\right) }=\left( -1\right) ^{n}%
\frac{n!\left( n+1\right) !}{\left( 2n+1\right) !}\left\langle \left( z^{-1}%
\text{d}z\right) ^{2n+1}\right\rangle\,.
\end{equation}%
Thus, by virtue of eq. (\ref{key}), it is possible to write down the
transgression form corresponding to $A$ and $A^{z}$ in terms of the pure
gauge connection and a total derivative%
\begin{equation}
Q_{A^{z}\leftarrow A}^{\left( 2n+1\right) }=Q_{z^{-1}\text{d}z\leftarrow
0}^{\left( 2n+1\right) }+\text{d}\left( \alpha _{2n}\left( A,z\right)
-Q_{A^{z}\leftarrow A\leftarrow 0}^{\left( 2n\right) }\right)\,.  \label{Q2}
\end{equation}%
Given a gauge group, the so-called gWZW action is defined as the boundary
action that appears from the transgression action in accordance with the
Stoke's theorem 
\begin{eqnarray}
S_{\text{gWZW}}\left[ A\right]  &=&\kappa \int_{M}Q_{A^{z}\leftarrow
A}^{\left( 2n+1\right) }  \notag \\
&=&\kappa \int_{\partial M}\alpha _{2n}\left( A,z\right) -Q_{A^{z}\leftarrow
A\leftarrow 0}^{\left( 2n\right) }\,.
\end{eqnarray}%
Since the transgression Lagrangian is odd-dimensional, the gWZW action
principle is always even-dimensional. As it happens in the SW-GN formalism,
the zero-forms $\xi $ are not longer interpreted as the parameters of a
symmetry transformation but as physical fields with a topological origin.
However, in contrast with the gauge invariant action principles that are
obtained in the SW-NG formalism, gWZW action principles are not exclusively
functionals of the non-linear gauge fields, but of the linear ones and the
zero-form multiplets.

\section{$\mathcal{N}=1$ Poincar\'{e} supergravity}\label{sec3}

\subsection{Chern--Simons supergravity}

The construction of a four-dimensional gWZW model requires the
five-dimensional CS action as starting point. We first consider
the $\mathcal{N}=1$ supersymmetric extension of the five-dimensional Poincar%
\'{e} algebra $\mathfrak{u}\left( 4|1\right) $, which is spanned by the set
of generators $\left\{ J_{AB},P_{A},K,Q^{\alpha },\bar{Q}_{\alpha }\right\} $
where $\bar{Q}_{\alpha }$ and $Q^{\alpha }$ are independent Dirac spinors.
Capital latin letters denote five-dimensional Lorentz indices taking values
as $A=0,\ldots 4$, while Greek letters denote spinor indices taking values
as $\alpha =1,\ldots 4$. In the chosen basis, the (anti)conmutation
relations between the introduced generators are given by \cite%
{Izaurieta:2005vp,Izaurieta:2006wv,Banados:1996hi}%
\begin{eqnarray}
\left[ J_{AB},P_{C}\right] &=&\eta _{BC}P_{A}-\eta _{AC}P_{B}\,, \notag \\
\left[ J_{AB},J_{CD}\right] &=&\eta _{BC}J_{AD}+\eta _{AD}J_{BC}-\eta
_{AC}J_{BD}-\eta _{BD}J_{AC}\,, \notag \\
\left[ J_{AB},Q^{\alpha }\right] &=&-\frac{1}{2}\left( \Gamma _{AB}\right)
_{~~\beta }^{\alpha }Q^{\beta }\,, \notag \\
\left[ J_{AB},\bar{Q}_{\alpha }\right] &=&\frac{1}{2}\left( \Gamma
_{AB}\right) _{~~\alpha }^{\beta }\bar{Q}_{\beta }\,, \notag \\
\left\{ Q^{\alpha },\bar{Q}_{\beta }\right\} &=&2\left( \Gamma ^{A}\right)
_{~~\beta }^{\alpha }P_{A}-4i\delta _{\beta }^{\alpha }K\,, \label{SP}
\end{eqnarray}%
where the metric signature is chosen as $\eta ^{AB}=\mathrm{diag}\left(
-,+,+,+,+\right) $. This superalgebra allows an inner invariant rank-$3$
product with the following components:%
\begin{eqnarray}
\left\langle KJ_{AB}J_{CD}\right\rangle &=&-\frac{i}{4}\left( \eta _{AC}\eta
_{BD}-\eta _{BC}\eta _{AD}\right)\,,  \notag \\
\left\langle J_{AB}J_{CD}P_{E}\right\rangle &=&\frac{1}{2}\epsilon _{ABCDE}\,,
\notag \\
\left\langle Q^{\alpha }J_{AB}\bar{Q}_{\beta }\right\rangle &=&-\left(
\Gamma _{AB}\right) _{\text{ \ }\beta }^{\alpha }\,.  \label{IT}
\end{eqnarray}%
We gauge the algebra by considering a one-form gauge connection $A$ with
non-vanishing gauge curvature $F=$d$A+\frac{1}{2}[A,A]$, to whose components
we denote 
\begin{eqnarray}
A &=&h^{A}P_{A}+\frac{1}{2}\omega ^{AB}J_{AB}+bK+\bar{\psi}_{\alpha
}Q^{\alpha }-\bar{Q}_{\alpha }\psi ^{\alpha }, \notag \\
F &=&\mathcal{T}^{A}P_{A}+\frac{1}{2}\mathcal{R}^{AB}J_{AB}+F_{b}K+\mathcal{%
\bar{F}}_{\alpha }Q^{\alpha }-\bar{Q}_{\alpha }\mathcal{F}^{\alpha },
\end{eqnarray}%
The components of the gauge curvature are given explicitly by 
\begin{eqnarray}
\mathcal{T}^{A} &=&\text{d}h^{A}+\omega _{\text{ \ }C}^{A}h^{C}-2\bar{\psi}%
_{\alpha }\left( \Gamma ^{A}\right) _{~~\beta }^{\alpha }\psi ^{\alpha }\,, \notag \\
\mathcal{R}^{AB} &=&\text{d}\omega ^{AB}+\omega _{\text{ \ }C}^{A}\omega
^{CB}\,, \notag \\
F_{b} &=&\text{d}b+4i\delta _{\beta }^{\alpha }\bar{\psi}_{\alpha }\psi
^{\alpha }\,, \notag \\
\mathcal{\bar{F}}_{\alpha } &=&\mathcal{D}\bar{\psi}_{\alpha }\equiv \text{d}%
\bar{\psi}_{\alpha }-\frac{1}{4}\omega ^{AB}\bar{\psi}_{\beta }\left( \Gamma
_{AB}\right) _{\text{ \ }\alpha }^{\beta }\,, \notag \\
\mathcal{F}^{\alpha } &=&\mathcal{D}\psi \equiv \text{d}\psi ^{\alpha }+%
\frac{1}{4}\omega ^{AB}\left( \Gamma _{AB}\right) _{\text{ \ }\beta
}^{\alpha }\psi ^{\beta }\,,
\end{eqnarray}%
where $\mathcal{D}$ denotes the covariant derivative defined with respect to
the five-dimensional spin connection $\omega ^{AB}$. By using the subspace
separation procedure (see refs. \cite{Izaurieta:2005vp,Izaurieta:2006wv}),
it is possible to write down the five-dimensional CS Lagrangian
in a convenient way:%
\begin{equation}
\mathcal{L}_{\text{CS}}\left( A\right) =\kappa \left( \frac{1}{4}\epsilon
_{ABCDE}\mathcal{R}^{AB}\mathcal{R}^{CD}h^{E}+\frac{i}{4}\mathcal{R}^{AB}%
\mathcal{R}_{AB}b-\left( \bar{\psi}\mathcal{R}^{AB}\Gamma _{AB}\mathcal{D}%
\psi +\mathcal{D}\bar{\psi}\mathcal{R}^{AB}\Gamma _{AB}\psi \right) \right)\,,
\end{equation}%
where $\kappa $ is a constant.

\subsection{Non-linear realization}

In order to perform the dimensional reduction, it is necessary to introduce
a non-linear realization of the gauge supergroup. We therefore consider a
second gauge connection $A^{z}$ related with $A$ by means of the following
large gauge transformation%
\begin{equation}
A^{z}=z^{-1}\left( \text{d}+A\right) z.  \label{Az}
\end{equation}%
Here, $z$ is an element of the gauge group, specifically in the coset space $%
\mathcal{G}_{5}/SO\left( 4,1\right) $. Taking in account the following
decomposition of the Poincar\'{e} superalgebra%
\begin{eqnarray}
L_{0} &=&\left\{ J_{AB}\right\} ,\text{ \ \ }L_{1}=\left\{
J_{AB},P_{A}\right\} ,\text{ \ \ }L_{3}=\left\{ J_{AB},P_{A},K\right\}\,, \notag \\
L_{4} &=&\left\{ J_{AB},P_{A},K,\bar{Q}_{\alpha }\right\} ,\text{ \ \ }%
L_{5}=\left\{ J_{AB},P_{A},K,\bar{Q}_{\alpha }\right\}\,,
\end{eqnarray}%
we can express $A^{z}$ \ by using the following gauge group element%
\begin{equation}
z=z_{\bar{\chi}}z_{\chi }z_{\varphi }z_{\phi }=e^{-\bar{\chi}_{\alpha
}Q^{\alpha }}e^{\bar{Q}_{\beta }\chi ^{\beta }}e^{-\varphi K}e^{-\phi
^{A}P_{A}}\,,  \label{vi}
\end{equation}%
where $\varphi $ and $\phi ^{A}$ are zero-forms, and where $\chi $ and $\bar{%
\chi}_{\alpha }$ are Dirac spinors zero-forms. In order to explicitly write
down for $A^{z}$ in terms of the components of $A$ and the parameters of the
gauge transformation, we use the following identities \cite{Zumino:1977av}%
\begin{eqnarray}
e^{X}\delta e^{-X} &=&\frac{\left( 1-e^{x}\right) }{X}\wedge \delta X\,,
\notag \\
e^{X}Ye^{-X} &=&e^{X}\wedge Y\,,  \label{id2}
\end{eqnarray}%
with the notation $X\wedge Y=\left[ X,Y\right] $ and where $\delta $ is any
variation. By directly and successively applying in the (anti)commutation
relations of the gauge algebra into eq. (\ref{Az}), we obtain the following
transformed gauge field%
\begin{equation}
A^{z}=V+W+\tilde{B}+\bar{\Psi}-\Psi\,,
\end{equation}%
where each component is given by%
\begin{eqnarray}
V &=&V^{A}P_{A}=\left( h^{A}-\mathcal{D}\phi ^{A}+2\bar{\psi}\Gamma ^{A}\chi
-2\mathcal{D}\bar{\chi}\Gamma ^{A}\chi -2\bar{\chi}\Gamma ^{A}\psi \right)
P_{A}\,,  \notag \\
W &=&\frac{1}{2}W^{AB}J_{AB}=\frac{1}{2}\omega ^{AB}J_{AB},  \notag \\
\tilde{B} &=&BK=\left\{ b-\text{d}\varphi +4i\left( \left( \mathcal{D}\bar{%
\chi}\right) \chi -\bar{\psi}\chi +\bar{\chi}\psi \right) \right\} K\,,
\notag \\
\bar{\Psi} &=&\bar{\Psi}Q=\left( \bar{\psi}-\mathcal{D}\bar{\chi}\right) Q\,,
\notag \\
\Psi  &=&\bar{Q}\Psi =\bar{Q}\left( \psi -\mathcal{D}\chi \right)\,.
\label{psi2}
\end{eqnarray}%
The non-linear realization of the gauge algebra allows the construction of
invariant action principles. In fact, the five-dimensional standard
supergravity theory, whose Lagrangian functional includes the
Einstein--Hilbert and Rarita--Schwinger terms becomes invariant under the
Poincar\'{e} superalgebra when one identifies the non-linear field $V^{A}$
as the f\"{u}nfbein field associated to the metric tensor of the
supergravity theory, and $\Psi ^{\alpha }$ and $\bar{\Psi}_{\alpha }$ as the
spin $3/2$ fields.

\section{Four-dimensional Poincar\'{e} supergravity}\label{sec4}

In order to write down the transgression form depending on both gauge
connections $A$ and $A^{z}$, let us recall (\ref{vi}). By inspection of the
(anti)commutation relations and invariant tensors of the gauge algebra, it
follows that, in this case, the pure gauge connection $z^{-1}$d$z$ has no
components along the Lorentz rotation generators. Therefore, the pure gauge
contribution to the transgression form, vanishes for any gauge parameter
lying in $\mathfrak{u}\left( 4|1\right) /\mathfrak{so}\left( 4,1\right) $%
\begin{equation}
Q_{z^{-1}\text{d}z\leftarrow 0}^{\left( 5\right) }=0\,,
\end{equation}%
As a consequence, the transgression form $Q_{A^{z}\leftarrow A}^{\left(
5\right) }$ is always exact and, according eq. (\ref{Q2}), can be written as 
\begin{equation}
Q_{A^{z}\leftarrow A}^{\left( 5\right) }=\text{d}\left( \alpha _{4}\left(
A,z\right) -Q_{A^{z}\leftarrow A\leftarrow 0}^{\left( 4\right) }\right)\,.
\label{tr5}
\end{equation}%
To find an explicit expression of this transgression, let us first consider
eq. (\ref{key}) with the following choice of gauge connections:%
\begin{equation}
Q_{A^{z}\leftarrow A}^{\left( 5\right) }=Q_{A^{z}\leftarrow \omega }^{\left(
5\right) }-Q_{A\leftarrow \omega }^{\left( 5\right) }-\text{d}%
Q_{A^{z}\leftarrow A\leftarrow \omega }^{\left( 4\right) }\,,  \label{trb}
\end{equation}%
Notice that we now choose the intermediate connection as $\bar{A}=\omega $.
A direct calculation shows that the difference between both transgression
forms in the r.h.s. of eq. (\ref{trb}) is given by%
\begin{align}
Q_{A^{z}\leftarrow \bar{A}}^{\left( 5\right) }&- Q_{A\leftarrow \bar{A}%
}^{\left( 5\right) } =\text{d}\left( \frac{3}{8}\epsilon _{ABCDE}\mathcal{R}%
^{AB}\mathcal{R}^{CD}\phi ^{E}+\frac{3i}{8}\mathcal{R}^{AB}\mathcal{R}%
_{AB}\varphi \right.  \left. +3i\left\langle \mathcal{D}\bar{\chi}\mathcal{R}\psi -\bar{\psi}%
\mathcal{RD}\chi +\mathcal{D}\bar{\chi}\mathcal{RD}\chi \right\rangle
\right) \notag \\
&  -\frac{3}{4}\epsilon _{ABCDE}\mathcal{R}^{AB}\mathcal{R}^{CD}\left( 
\bar{\psi}\Gamma ^{E}\chi -\mathcal{D}\bar{\chi}\Gamma ^{E}\chi -\bar{\chi}%
\Gamma ^{E}\psi \right)  +\frac{3}{2\ell }\mathcal{R}^{AB}\mathcal{R}_{AB}\left( \left( 
\mathcal{D}\bar{\chi}\right) \chi -\bar{\psi}\chi +\bar{\chi}\psi \right) \notag \\
& -3\left( \mathcal{D}^{2}\bar{\chi}\mathcal{R}^{AB}\Gamma _{AB}\psi +\bar{%
\psi}\mathcal{R}^{AB}\Gamma _{AB}\mathcal{D}^{2}\chi -\mathcal{D}\bar{\chi}%
\mathcal{R}^{AB}\Gamma _{AB}\mathcal{D}^{2}\chi \right)\,,   
\end{align}%
where letters carrying no index inside the trace are vectors of the Lie
superalgebra, i.e., $\mathcal{R=}\frac{1}{2}\mathcal{R}^{AB}J_{AB}$, $\psi =%
\bar{Q}_{\alpha }\psi ^{\alpha }$, $\bar{\psi}=\bar{\psi}_{\alpha }Q^{\alpha
}$, $\chi =\bar{Q}_{\alpha }\chi ^{\alpha }$ and $\bar{\chi}=\bar{\chi}%
_{\alpha }Q^{\alpha }$. By using the Bianchi identities, we have that the
second derivatives of the fermion zero-forms can be written in terms of the
Lorentz curvature as 
\begin{align}
\mathcal{D}^{2}\bar{\chi}_{\alpha }&=-\frac{1}{4}\bar{\chi}_{\beta }\mathcal{R%
}^{AB}\left( \Gamma _{AB}\right) _{\text{ \ }\alpha }^{\beta }\,, &\mathcal{D}^{2}\chi ^{\alpha }&=\frac{1}{4}\mathcal{R}^{AB}\left( \Gamma
_{AB}\right) _{\text{ \ }\beta }^{\alpha }\chi ^{\beta }\,.
\end{align}%
Then, by integrating by parts and using the properties of the gamma matrices
in five dimensions, we find that the difference between both transgression
forms is given by the following total derivative%
\begin{equation}
Q_{A^{z}\leftarrow \bar{A}}^{\left( 5\right) }-Q_{A\leftarrow \bar{A}%
}^{\left( 5\right) }=\text{d}\left[ \frac{3}{8}\epsilon _{ABCDE}\mathcal{R}%
^{AB}\mathcal{R}^{CD}\phi ^{E}+\frac{3i}{8}\mathcal{R}^{AB}\mathcal{R}%
_{AB}\varphi +3i\left\langle \mathcal{D}\bar{\chi}\mathcal{R}\psi -\bar{\psi}%
\mathcal{RD}\chi +\mathcal{D}\bar{\chi}\mathcal{RD}\chi \right\rangle \right]\,.  \label{dif2}
\end{equation}%
On the other hand, the the boundary term in the r.h.s. of eq. (\ref{trb})
can be obtained from eq. (\ref{bound}) by setting $n=2$, $A_{2}=A^{z}$, $%
A_{1}=A$ and $\bar{A}=\omega $. Consequently, the homotopic gauge field
becomes $A_{st}=\omega +s\left( A^{z}-A\right) +t\left( A-\omega \right) $.
A direct integration leads to%
\begin{equation}
Q_{A^{z}\leftarrow A\leftarrow \bar{A}}^{\left( 4\right) }=-3i\left\langle 
\mathcal{D}\bar{\chi}\mathcal{R}\psi -\bar{\psi}\mathcal{RD}\chi
\right\rangle\,.  \label{q4}
\end{equation}%
Then, by plugging in eqs. (\ref{dif2}) and (\ref{q4}) into (\ref{trb}), we
finally obtain an explicit expression for the transgression form in terms of
a total derivative%
\begin{eqnarray}
Q_{A^{z}\leftarrow A}^{\left( 5\right) } &=&\text{d}\left\{ \frac{3}{8}%
\epsilon _{ABCDE}\mathcal{R}^{AB}\mathcal{R}^{CD}\phi ^{E}+\frac{3i}{8}%
\mathcal{R}^{AB}\mathcal{R}_{AB}\varphi \right.   \notag \\
&&\left. +3\left[ \mathcal{D}\bar{\chi}\mathcal{R}^{AB}\Gamma _{AB}\left(
\psi +\frac{1}{4}\mathcal{D}\chi \right) -\left( \bar{\psi}-\frac{1}{4}%
\mathcal{D}\bar{\chi}\right) \mathcal{R}^{AB}\Gamma _{AB}\mathcal{D}\chi %
\right] \right\}\,.
\end{eqnarray}%
Therefore, the four-dimensional induced action is given by%
\begin{eqnarray}
S &=&\kappa \int \left[ \epsilon _{ABCDE}\mathcal{R}^{AB}\mathcal{R}%
^{CD}\phi ^{E}+i\mathcal{R}^{AB}\mathcal{R}_{AB}\varphi \right.   \notag \\
&&\left. -8\left( \bar{\chi}\mathcal{R}^{AB}\Gamma _{AB}\mathcal{D}\psi +%
\mathcal{D}\bar{\psi}\mathcal{R}^{AB}\Gamma _{AB}\chi -\frac{1}{2}\mathcal{D}%
\bar{\chi}\mathcal{R}^{AB}\Gamma _{AB}\mathcal{D}\chi \right) \right]\,.
\label{action4}
\end{eqnarray}%
This action in analogue to the one found in ref. \cite{Salgado:2014jka} for $%
\left( 1+1\right) $-dimensional supergravity. It is invariant under the
transformations of the five-dimensional Poincar\'{e} supergroup and it can
be interpreted as a supersymmetric extension of the topological gravity
introduced in ref. \cite{Chamseddine:1990gk}, and alternatively found as a
gWZW action in ref. \cite{Salgado:2013pva} for the bosonic case.

\subsection{Decomposition of the action}

The obtained gWZW action from eq. (\ref{action4}) is four-dimensional.
However, it is a functional of the original five-dimensional field content
of the transgression field theory. When gauging Poincar\'{e} or AdS
supergroups, the gauge field associated to the translation operator is
usually identified as the vielbein of the corresponding supergravity theory,
and therefore it is considered that it carries the information about the
metric in the resulting field equations. Notice that, at this point, we have
not yet introduced a notion of metricity in the supergravity theory.
Moreover, the field $h^{A}$ has been removed from the functional in the
dimensional reduction process and it is not longer present in the action
principle in eq. (\ref{action4}). This is a common feature of gWZW models
originated in the gauging of space-time symmetries, and allows us to
identify the some components of the five-dimensional spin connection as
vierbein in a four-dimensional supergravity theory. Thus, the original gauge
invariance under the five-dimensional Lorentz group is now interpreted as
invariance under the four-dimensional de Sitter group. We therefore
decompose the index $A=\left( a,4\right) $ with $a=0,1,2,3$, and rename $%
\omega ^{a4}=-\omega ^{4a}=e^{a}$ as vierbein one-form. The five-dimensional
Lorentz curvature is also decomposed, as follows: 
\begin{eqnarray}
\mathcal{R}^{ab} &=&\text{d}\omega ^{ab}+\omega _{\text{ \ }c}^{a}\omega
^{cb}+\omega _{\text{ \ }4}^{a}\omega ^{4b}=R^{ab}-e^{a}e^{b}\,, \notag \\
\mathcal{R}^{a4} &=&\text{D}e^{a}=T^{a}\,.
\end{eqnarray}%
Consequently, we split the action principle into its bosonic and fermionic
sectors as $S=S_{\text{B}}+S_{\text{F}}$, being $S_{\text{F}}$ the
contribution depending on spinor fields in the r.h.s. of eq. (\ref{action4}%
). In terms of the previous decomposition, the bosonic sector of the action
is given by%
\begin{eqnarray}
S_{\text{B}} &=&\kappa \int \left[ -4\left( \epsilon _{abcd}R^{ab}e^{c}-%
\frac{1}{3}\epsilon _{abcd}e^{a}e^{b}e^{c}\right) \text{D}\phi ^{d}\right. 
\notag \\
&&\left. +\epsilon _{abcd}\left[
R^{ab}R^{cd}-2R^{ab}e^{c}e^{d}+e^{a}e^{b}e^{c}e^{d}\right] \phi
^{4}+iR^{ab}R_{ab}\varphi +2iT^{a}e_{a}\text{d}\varphi \right]\,.
\end{eqnarray}%
In the same way, the fermionic sector of the action is given in terms of the
four-dimensional quantities as follows%
\begin{eqnarray}
S_{\text{F}} &=&\kappa \int -8\left[ \bar{\chi}\left(
R^{ab}-e^{a}e^{b}\right) \Gamma _{ab}\mathcal{D}\psi +2\bar{\chi}T^{a}\Gamma
_{a}\Gamma \mathcal{D}\psi +\mathcal{D}\bar{\psi}\left(
R^{ab}-e^{a}e^{b}\right) \Gamma _{ab}\chi \right.  \notag \\
&&\left. -\frac{1}{2}\mathcal{D}\bar{\chi}\left( R^{ab}-e^{a}e^{b}\right)
\Gamma _{ab}\mathcal{D}\chi -\mathcal{D}\bar{\chi}T^{a}\Gamma _{a}\Gamma 
\mathcal{D}\chi \right]\,,
\end{eqnarray}%
where the five-dimensional Lorentz covariant derivatives are given in terms
of the four-dimensional ones according to\footnote{%
From now on, we denote $\Gamma ^{5}\equiv \Gamma $.}%
\begin{equation}
\mathcal{D}\bar{\psi}=\text{D}\bar{\psi}-\frac{1}{2}e^{a}\bar{\psi}\Gamma
_{a}\Gamma ,\text{ \ \ \ \ }\mathcal{D}\psi =\text{D}\psi +\frac{1}{2}%
e^{a}\Gamma _{a}\Gamma \psi\,.
\end{equation}

\section{Dynamics}\label{sec5}

From now on, we will denote as $\mathcal{L}_{\text{G}}$ to the bosonic
Lagrangian four-form, and identify to the fermionic contribution as a matter
Lagrangian, i.e.,%
\begin{eqnarray}
\mathcal{L}_{\text{G}} &=&\kappa \left[ \epsilon _{ABCDE}\mathcal{R}^{AB}%
\mathcal{R}^{CD}\phi ^{E}+i\mathcal{R}^{AB}\mathcal{R}_{AB}\varphi \right]\,, \notag
\\
\mathcal{L}_{\text{M}} &=&-8\kappa \left[ \bar{\chi}\mathcal{R}^{AB}\Gamma
_{AB}\mathcal{D}\psi +\mathcal{D}\bar{\psi}\mathcal{R}^{AB}\Gamma _{AB}\chi -%
\frac{1}{2}\mathcal{D}\bar{\chi}\mathcal{R}^{AB}\Gamma _{AB}\mathcal{D}\chi %
\right]\,.
\end{eqnarray}%
Although we hold the writing in terms of the five-dimensional indices for
convenience, it is important to recall that they describe a four-dimensional
theory with $\omega ^{AB}$ packing the spin connection and vielbein forms,
while $\mathcal{R}^{AB}$ contains the Lorentz curvature and torsion. In
these terms, we introduce a generalized spin form $\Sigma _{AB}$, such that
the variation of $\mathcal{L}_{\text{M}}$ with respect to $\omega ^{AB}$ is
given by%
\begin{equation}
\delta _{\omega }\mathcal{L}_{\text{M}}=-k\delta \omega ^{AB}\ast \Sigma
_{AB}\,,
\end{equation}%
with $\ast $ the Hodge dual operator and $k$ a dimensional constant. The
components of the generalized spin form are split into the four dimensional
spin form and energy-momentum forms, as follows%
\begin{equation}
\Sigma _{a4}=\mathcal{T}_{a},\text{ \ \ \ \ }\Sigma _{AB}=\sigma _{ab}\,.
\end{equation}%
Therefore, the field equations can be written as%
\begin{equation}
\delta _{\omega }\mathcal{L}_{\text{G}}-k\delta \omega ^{AB}\ast \Sigma
_{AB}=0\,.
\end{equation}%
The variation of $\mathcal{L}_{\text{G}}$ with respect to $\omega ^{AB}$ is
given by%
\begin{equation}
\delta _{\omega }\mathcal{L}_{\text{G}}=2\kappa \delta \omega ^{AB}\left[
\epsilon _{ABCDE}\mathcal{R}^{CD}\mathcal{D}\phi ^{E}+i\mathcal{R}_{AB}\text{%
d}\varphi \right]\,.
\end{equation}%
On the other hand, the field variation of the matter Lagrangian is given by%
\begin{eqnarray}
\delta _{\omega }\mathcal{L}_{\text{M}} &=&-8\kappa \left[ -\mathcal{D}\bar{%
\chi}\delta \omega ^{AB}\Gamma _{AB}\mathcal{D}\psi +\bar{\chi}\delta \omega
^{AB}\Gamma _{AB}\mathcal{D}^{2}\psi +\frac{1}{4}\bar{\chi}\mathcal{R}%
^{AB}\delta \omega ^{CD}\Gamma _{AB}\Gamma _{CD}\psi \right.   \notag \\
&&-\frac{1}{4}\delta \omega ^{CD}\bar{\psi}\mathcal{R}^{AB}\Gamma
_{CD}\Gamma _{AB}\chi -\mathcal{D}^{2}\bar{\psi}\delta \omega ^{AB}\Gamma
_{AB}\chi +\mathcal{D}\bar{\psi}\delta \omega ^{AB}\Gamma _{AB}\mathcal{D}%
\chi    \notag \\
&&+\frac{1}{8}\delta \omega ^{CD}\bar{\chi}\mathcal{R}^{AB}\Gamma
_{CD}\Gamma _{AB}\mathcal{D}\chi-\frac{1}{8}\mathcal{D}\bar{\chi}\mathcal{R}^{AB}\delta \omega ^{CD}\Gamma
_{AB}\Gamma _{CD}\chi -\frac{1}{2}\mathcal{D}^{2}\bar{\chi}\delta \omega
^{AB}\Gamma _{AB}\mathcal{D}\chi \notag \\
&&-\frac{1}{2}\mathcal{D}\bar{\chi}\delta
\omega ^{AB}\Gamma _{AB}\mathcal{D}^{2}\chi\,,
\end{eqnarray}%
where we have used the identities%
\begin{eqnarray}
\delta \mathcal{D}\bar{\psi} &=&-\frac{1}{4}\delta \omega ^{AB}\bar{\psi}%
\Gamma _{AB},\text{ \ \ \ }\delta \mathcal{D}\psi =\frac{1}{4}\delta \omega
^{AB}\Gamma _{AB}\psi\,, \notag \\
\delta \mathcal{D}\bar{\chi} &=&-\frac{1}{4}\delta \omega ^{AB}\bar{\chi}%
\Gamma _{AB},\text{ \ \ \ }\delta \mathcal{D}\chi =\frac{1}{4}\delta \omega
^{AB}\Gamma _{AB}\chi\,.
\end{eqnarray}%
By integrating by parts and plugging in the Bianchi identities%
\begin{eqnarray}
\mathcal{D}^{2}\bar{\chi} &=&-\frac{1}{4}\mathcal{R}^{AB}\bar{\chi}\Gamma
_{AB},\text{ \ \ \ }\mathcal{D}^{2}\chi =\frac{1}{4}\mathcal{R}^{AB}\Gamma
_{AB}\chi\,, \notag\\
\mathcal{D}^{2}\bar{\psi} &=&-\frac{1}{4}\mathcal{R}^{AB}\bar{\psi}\Gamma
_{AB},\text{ \ \ \ }\mathcal{D}^{2}\psi =\frac{1}{4}\mathcal{R}^{AB}\Gamma
_{AB}\psi\,,
\end{eqnarray}%
we obtain%
\begin{eqnarray}
\delta _{\omega }\mathcal{L}_{\text{M}} &=&-8\kappa \delta \omega ^{AB}\left[
\mathcal{D}\bar{\chi}\Gamma _{AB}\mathcal{D}\psi +\mathcal{D}\bar{\psi}%
\Gamma _{AB}\mathcal{D}\chi +\mathcal{R}_{AB}\left( \bar{\psi}\chi -\bar{\chi%
}\psi -\frac{1}{2}\text{d}\left( \bar{\chi}\chi \right) \right) \right.  
\notag \\
&&\left. +\frac{1}{2}\epsilon _{ABCDE}\mathcal{R}^{CD}\left( \bar{\psi}%
\Gamma ^{E}\chi -\bar{\chi}\Gamma ^{E}\psi -\frac{1}{2}\mathcal{D}\left( 
\bar{\chi}\Gamma ^{E}\chi \right) \right) \right]\,.
\end{eqnarray}%
Finally we have an expression for the dual spin form%
\begin{eqnarray}
\ast \Sigma _{AB} &=&\frac{8\kappa }{k}\left[ \mathcal{D}\bar{\chi}\Gamma
_{AB}\mathcal{D}\psi +\mathcal{D}\bar{\psi}\Gamma _{AB}\mathcal{D}\chi +%
\mathcal{R}_{AB}\left( \bar{\psi}\chi -\bar{\chi}\psi -\frac{1}{2}\text{d}%
\left( \bar{\chi}\chi \right) \right) \right.   \notag \\
&&\left. +\frac{1}{2}\epsilon _{ABCDE}\mathcal{R}^{CD}\left( \bar{\psi}%
\Gamma ^{E}\chi -\bar{\chi}\Gamma ^{E}\psi -\frac{1}{2}\mathcal{D}\left( 
\bar{\chi}\Gamma ^{E}\chi \right) \right) \right]\,.
\end{eqnarray}%
The field equations coming from the variation of the action with respect to $%
\omega ^{AB}$ are therefore given by $\varepsilon _{AB}=0$ with%
\begin{eqnarray}
\varepsilon _{AB} &=&\epsilon _{ABCDE}\mathcal{R}^{CD}\mathcal{D}\phi ^{E}+i%
\mathcal{R}_{AB}d\varphi -4\left[ \mathcal{D}\bar{\chi}\Gamma _{AB}\mathcal{D%
}\psi +\mathcal{D}\bar{\psi}\Gamma _{AB}\mathcal{D}\chi \right.   \notag \\
&&\left. +\mathcal{R}_{AB}\left( \bar{\psi}\chi -\bar{\chi}\psi -\frac{1}{2}%
\text{d}\left( \bar{\chi}\chi \right) \right) +\frac{1}{2}\epsilon _{ABCDE}%
\mathcal{R}^{CD}\left( \bar{\psi}\Gamma ^{E}\chi -\bar{\chi}\Gamma ^{E}\psi -%
\frac{1}{2}\mathcal{D}\left( \bar{\chi}\Gamma ^{E}\chi \right) \right) %
\right]\,. \notag\\
\end{eqnarray}%
or, equivalently in components, the field equations related with the
independent variations $\delta e^{a}$ and $\delta \omega ^{ab}$ are given by 
\begin{eqnarray}
\varepsilon _{a4} &=&-\epsilon _{abcd}\mathcal{R}^{bc}\left( D\phi
^{d}+e^{d}\phi ^{4}\right) +iT_{a}\text{d}\varphi   \notag \\
&&-4\left[ \mathcal{D}\bar{\chi}\Gamma _{a}\Gamma \mathcal{D}\psi +\mathcal{D%
}\bar{\psi}\Gamma _{a}\Gamma \mathcal{D}\chi +T_{a}\left( \bar{\psi}\chi -%
\bar{\chi}\psi -\frac{1}{2}\text{d}\left( \bar{\chi}\chi \right) \right)
\right.   \notag \\
&&\left. -\frac{1}{2}\epsilon _{abcd}\mathcal{R}^{bc}\left( \bar{\psi}\Gamma
^{d}\chi -\bar{\chi}\Gamma ^{d}\psi -\frac{1}{2}\text{D}\left( \bar{\chi}%
\Gamma ^{d}\chi \right) -\frac{1}{2}e^{d}\bar{\chi}\Gamma ^{4}\chi \right) %
\right] \,,
\end{eqnarray}%
\begin{eqnarray}
\varepsilon _{ab} &=&\epsilon _{abcd4}\mathcal{R}^{cd}\left( \text{d}\phi
^{4}-e^{b}\phi _{b}\right) -2\epsilon _{abcd}T^{c}\left( \text{D}\phi
^{d}+e^{d}\phi ^{4}\right) +i\mathcal{R}_{ab}\text{d}\varphi   \notag \\
&&-4\left[ \mathcal{D}\bar{\chi}\Gamma _{ab}\mathcal{D}\psi +\mathcal{D}\bar{%
\psi}\Gamma _{ab}\mathcal{D}\chi +\mathcal{R}_{ab}\left( \bar{\psi}\chi -%
\bar{\chi}\psi -\frac{1}{2}\text{d}\left( \bar{\chi}\chi \right) \right)
\right.   \notag \\
&&+\frac{1}{2}\epsilon _{abcd}\mathcal{R}^{cd}\left( \bar{\psi}\Gamma \chi -%
\bar{\chi}\Gamma \psi -\frac{1}{2}\text{d}\left( \bar{\chi}\Gamma \chi
\right) +\frac{1}{2}e_{b}\bar{\chi}\Gamma ^{b}\chi \right)   \notag \\
&&\left. -\epsilon _{abcd}T^{c}\left( \bar{\psi}\Gamma ^{d}\chi -\bar{\chi}%
\Gamma ^{d}\psi -\frac{1}{2}\text{D}\left( \bar{\chi}\Gamma ^{d}\chi \right)
-\frac{1}{2}e^{d}\left( \bar{\chi}\Gamma \chi \right) \right) \right]\,,
\end{eqnarray}%
where $D$ denotes the covariant derivative defined with respect to the four
dimensional spin connection $\omega ^{ab}$.

\section{Non-relativistic limit}\label{sec6}

\subsection{Chern--Simons theory}

Let us now consider a non-relativistic contraction of the gauge algebra \eqref{SP}. We
split Lorentz index $A$ in the space-time components as $A=\left( 0,I\right) 
$ with $I=1,\ldots ,4$. Moreover, we rename and perform the following
rescaling on the Poincar\'{e} superalgebra generators as in \cite%
{Andringa:2010it,Gonzalez:2016xwo}%
\begin{align}
P_{0} &\longrightarrow H\, & P_{I}&\longrightarrow \lambda P_{I}\,, &
J_{0I}&\longrightarrow \lambda G_{I}\,, \notag\\
Q^{\alpha } &\longrightarrow \sqrt{\lambda }Q^{\alpha }\,, & \bar{Q}%
_{\alpha }&\longrightarrow \sqrt{\lambda }\bar{Q}_{\alpha }\,, &
K&\longrightarrow \lambda K\,.
\end{align}%
When taking the limit $\lambda \rightarrow \infty $, the superalgebra
(anti)commutation relations become%
\begin{eqnarray}
\left[ J_{IJ},P_{K}\right]  &=&\eta _{JK}P_{I}-\eta _{IK}P_{J}\,, \notag\\
\left[ G_{I},H\right]  &=&P_{I}\,, \notag\\
\left[ J_{IJ},J_{KL}\right]  &=&\eta _{JK}J_{IL}+\eta _{IL}J_{JK}-\eta
_{IK}J_{JL}-\eta _{JL}J_{IK}\,, \notag\\
\left[ J_{IJ},G_{K}\right]  &=&\eta _{JK}G_{I}-\eta _{IK}G_{J}\,, \notag\\
\left[ J_{IJ},Q^{\alpha }\right]  &=&-\frac{1}{2}\left( \Gamma _{IJ}\right)
_{~~\beta }^{\alpha }Q^{\beta }\,,\notag \\
\left[ J_{IJ},\bar{Q}_{\alpha }\right]  &=&\frac{1}{2}\left( \Gamma
_{IJ}\right) _{~~\alpha }^{\beta }\bar{Q}_{\beta }\,, \notag\\
\left\{ Q^{\alpha },\bar{Q}_{\beta }\right\}  &=&2\left( \Gamma ^{I}\right)
_{~~\beta }^{\alpha }P_{I}-4i\delta _{\beta }^{\alpha }K\,.
\end{eqnarray}%
The non-relativistic limit of the Poincaré superalgebra \eqref{SP} reproduces a supersymmetric extension of the Galilei algebra \cite{Bacry:1968zf}. We now introduce a one-form gauge connection $A$ and the corresponding gauge
curvature $F$, to whose components we denote%
\begin{eqnarray}
A &=&\tau H+h^{I}P_{I}+\omega ^{I}G_{I}+\frac{1}{2}\omega ^{IJ}J_{IJ}+bK+%
\bar{\psi}_{\alpha }Q^{\alpha }-\bar{Q}_{\alpha }\psi ^{\alpha }\,,
\label{Anr} \\
F &=&\hat{T}H+\hat{T}^{I}P_{I}+R^{I}G_{I}+\frac{1}{2}\mathcal{R}%
^{IJ}J_{IJ}+F_{b}K+\mathcal{\bar{F}}_{\alpha }Q^{\alpha }-\bar{Q}_{\alpha }%
\mathcal{F}^{\alpha }\,.  \label{Fnr}
\end{eqnarray}%
The components of the new gauge curvature are explicitly given by%
\begin{eqnarray}
\hat{T} &=&\text{d}\tau\,, \notag\\
\hat{T}^{I} &=&\text{D}_{\omega }h^{I}+\omega ^{I}\tau -2\bar{\psi}_{\alpha
}\left( \Gamma ^{I}\right) _{~~\beta }^{\alpha }\psi ^{\alpha }\,, \notag\\
\mathcal{R}^{I} &=&\text{D}_{\omega }\omega ^{I}\,, \notag\\
\mathcal{R}^{IJ} &=&\text{d}\omega ^{IJ}+\omega _{\text{ \ }K}^{I}\omega
^{KJ}\,, \notag\\
F_{b} &=&\text{d}b+4i\delta _{\beta }^{\alpha }\bar{\psi}_{\alpha }\psi
^{\alpha }\,, \notag\\
\mathcal{\bar{F}}_{\alpha } &=&\text{D}_{\omega }\bar{\psi}_{\alpha }, \notag\\
\mathcal{F}^{\alpha } &=&\text{D}_{\omega }\psi ^{\alpha }\,,
\end{eqnarray}%
where D$_{\omega }$ is the covariant derivative with respect to the spatial
spin connection $\omega ^{IJ}$. 

At the level of the invariant tensor, one can check that the non-relativistic limit of the non-vanishing components \eqref{IT} reproduces
\begin{align}
    \langle J_{IJ}J_{KL}H\rangle&= \frac{1}{2}\epsilon _{IJKL}\,,
\end{align}
with the convention $\epsilon _{0IJKL}=\epsilon _{IJKL}$. Then, the five-dimensional
CS Lagrangian takes the form%
\begin{equation}
\mathcal{L}_{\text{CS}}^{\text{NR}}=\frac{\kappa }{4}\epsilon _{IJKL}%
\mathcal{R}^{IJ}\mathcal{R}^{KL}\tau\,,
\end{equation}%
Note that,
although the non-relativistic limit of the Poincar\'{e} superalgebra is a
supersymmetric extension of the Galilei algebra, the CS
Lagrangian does not lead to supergravity. This is a consequence of the fact
that, in this limit, the invariant tensor of the algebra only carries
non-zero components in the bosonic sector. As we shall see in the next section, the Carrollian limit preserves supergravity. In a future work, it would be interesting to explore the existence of a non-relativistic counterpart of Poincaré supergravity that preserves supersymmetry.

\subsection{gWZW model}

Let us now consider the non-relativistic limit of the gWZW action principle
obtained in section \ref{sec4}. As we did in the relativistic case, in order to
construct the transgression field theory, we introduce a secondary gauge
connection $A^{z}$ related with $A$ by means of a gauge transformation. We
denote the components of $A^{z}$ and the supergroup parameters as follows:%
\begin{eqnarray}
A^{z} &=&\tau H+H^{I}P_{I}+W^{I}G_{I}+\frac{1}{2}W^{IJ}J_{IJ}+BK+\bar{\Psi}%
_{\alpha }Q^{\alpha }-\bar{Q}_{\alpha }\Psi ^{\alpha }\,,  \notag \\
z &=&e^{-\bar{\chi}_{\alpha }Q^{\alpha }}e^{\bar{Q}_{\beta }\chi ^{\beta
}}e^{-\varphi K}e^{-\phi ^{I}P_{I}}e^{-\phi H}\,.  \label{znr}
\end{eqnarray}%
The four dimensional action is reduced to%
\begin{equation}
\mathcal{L}_{\text{gWZW}}^{\text{NR}}=\kappa \epsilon _{IJKL}\mathcal{R}^{IJ}%
\mathcal{R}^{KL}\phi ,
\end{equation}%
where we rename $\phi ^{0}=\phi $. We now perform a second index
decomposition; the spatial index of the five-dimensional theory is split as $%
I=\left( i,4\right) $ where $i=1,2,3$ is the spatial index of the
non-relativistic four-dimensional theory. We also decompose the
non-relativistic spin connection and curvature as follows:%
\begin{eqnarray}
\omega ^{I} &=&\left( \omega ^{i},\tau \right)\,,  \notag \\
\omega ^{IJ} &=&\left( \omega ^{ij},\omega ^{i4}\right) \equiv \left( \omega
^{ij},e^{i}\right)\,,  \notag \\
\mathcal{R}^{I} &=&\left( \mathcal{R}^{i},T\right)\,, \notag \\
\mathcal{R}^{IJ} &=&\left( \mathcal{R}^{ij},\mathcal{R}^{i4}\right) \equiv
\left( \mathcal{R}^{ij},T^{i}\right)\,,  \label{RIJ}
\end{eqnarray}%
with%
\begin{eqnarray}
\mathcal{R}^{ij} &=&R^{ij}-e^{i}e^{j}\,,  \notag \\
T^{i} &=&\text{d}e^{i}+\omega _{\text{ \ }k}^{i}e^{k},  \label{Ti}
\end{eqnarray}%
where $R^{ij}=~$d$\omega ^{ij}+\omega _{\text{ \ }k}^{i}\omega ^{kj}$ is the 
$\mathfrak{so}\left( 3\right) $ curvature. Since $h^{A}$ is not longer
present in the theory, we interpret $e^{i}$ as spatial vielbein. In this
way, the Lagrangian density is written as%
\begin{equation}
\mathcal{L}_{\text{gWZW}}^{\text{NR}}=4\kappa \epsilon _{ijk}\mathcal{R}^{ij}%
\mathcal{R}^{k4}\phi =4\kappa \epsilon _{ijk}\left( R^{ij}-e^{i}e^{j}\right)
T^{k}\phi\,.
\end{equation}%
As it happens in the non-relativistic five-dimensional CS theory,
the resulting Lagrangian density is purely bosonic.

\section{Ultra-relativistic limit}\label{sec7}

\subsection{Chern--Simons theory}

Let us now consider the ultra relativistic limit of the five-dimensional
Poincar\'{e} superalgebra \cite{levy1965nouvelle,Bacry:1968zf,Duval:2014uoa}%
. With this purpose, we consider again the space-time splitting of the
Lorentz index $A$ in the Poincar\'{e} superalgebra. We rename and rescale
the generators as%
\begin{align}
P_{0} &\longrightarrow \lambda H\,,& J_{0I}&\longrightarrow \lambda
G_{I}\,, & K\longrightarrow \lambda K\,, \notag\\
Q^{\alpha } &\longrightarrow \sqrt{\lambda }Q^{\alpha }\,, & \bar{Q}%
_{\alpha }&\longrightarrow \sqrt{\lambda }\bar{Q}_{\alpha }\,.
\end{align}%
The resulting ultra-relativistic superalgebra is obtained by taking the
limit $\lambda \rightarrow \infty $, and is given by the following
supersymmetric extension of the Carroll algebra \cite{Bacry:1968zf} in five dimensions%
\begin{eqnarray}
\left[ J_{IJ},P_{K}\right]  &=&\eta _{JK}P_{I}-\eta _{IK}P_{J}\,, \notag \\
\left[ G_{I},P_{J}\right]  &=&\eta _{IJ}H\,, \notag \\
\left[ J_{IJ},J_{KL}\right]  &=&\eta _{JK}J_{IL}+\eta _{IL}J_{JK}-\eta
_{IK}J_{JL}-\eta _{JL}J_{IK}\,, \notag \\
\left[ J_{IJ},G_{K}\right]  &=&\eta _{JK}G_{I}-\eta _{IK}G_{J}\,, \notag \\
\left[ J_{IJ},Q^{\alpha }\right]  &=&-\frac{1}{2}\left( \Gamma _{IJ}\right)
_{~~\beta }^{\alpha }Q^{\beta }\,, \notag \\
\left[ J_{IJ},\bar{Q}_{\alpha }\right]  &=&\frac{1}{2}\left( \Gamma
_{IJ}\right) _{~~\alpha }^{\beta }\bar{Q}_{\beta }\,, \notag \\
\left\{ Q^{\alpha },\bar{Q}_{\beta }\right\}  &=&2\left( \Gamma ^{0}\right)
_{~~\beta }^{\alpha }H-4i\delta _{\beta }^{\alpha }K\,.
\end{eqnarray}%
Following the procedure used in the non-relativistic case, we now introduce a one-form gauge
connection $A$ and the corresponding gauge curvature $F=$d$A+A^{2}$. Since
the ultra-relativistic algebra has the same number of generators as its
non-relativistic analogue, we denote the components of $A$ and $F$ as they
are given in eqs. (\ref{Anr}) and (\ref{Fnr}) respectively. In this case,
the components of the gauge curvature are given by%
\begin{eqnarray}
\hat{T} &=&\text{d}\tau +\omega _{J}h^{J}-2\bar{\psi}_{\alpha }\left( \Gamma
^{0}\right) _{~~\beta }^{\alpha }\psi ^{\alpha }\,, \notag \\
\hat{T}^{I} &=&\text{D}_{\omega }h^{I}\,, \notag \\
\mathcal{R}^{IJ} &=&\text{d}\omega ^{IJ}+\omega _{\text{ \ }K}^{I}\omega
^{KJ}\,, \notag \\
\mathcal{R}^{I} &=&\text{D}_{\omega }\omega ^{I}\,, \notag \\
F_{b} &=&\text{d}b+4i\delta _{\beta }^{\alpha }\bar{\psi}_{\alpha }\psi
^{\alpha }\,, \notag \\
\mathcal{\bar{F}}_{\alpha } &=&\text{D}_{\omega }\bar{\psi}_{\alpha }\,, \notag \\
\mathcal{F}^{\alpha } &=&\text{D}_{\omega }\psi ^{\alpha }\,.
\end{eqnarray}%
The CS Lagrangian invariant under the transformation of this
algebra is given by%
\begin{eqnarray}
\mathcal{L}_{\text{CS}}^{\text{UR}} &=&\kappa \epsilon _{IJKL}\left( \frac{1%
}{4}\mathcal{R}^{IJ}\mathcal{R}^{KL}\tau +\mathcal{R}^{IJ}\mathcal{R}%
^{K}h^{L}\right)   \notag \\
&&+\frac{i\kappa }{4}\mathcal{R}^{IJ}\mathcal{R}_{IJ}b-\kappa \left( \bar{%
\psi}\mathcal{R}^{IJ}\Gamma _{IJ}\text{D}_{\omega }\psi +\text{D}_{\omega }%
\bar{\psi}\mathcal{R}^{IJ}\Gamma _{IJ}\psi \right)\,.
\end{eqnarray}%
In contrast with the non-relativistic Lagrangian, the invariant tensor still
carry non-vanishing components in the fermionic sector after taking the
limit and, as a consequence, the supergravity is preserved in the
ultra-relativistic regime.

\subsection{gWZW model}

Let us now consider the ultra-relativistic limit of the four dimensional
gWZW Lagrangian. As before, we introduce a group element $z$ and a
non-linear gauge field $A^{z}$, obtained from $A$ through a large gauge
transformation. We denote to the components of $z$ and $A^{z}$ as in eq. \eqref{znr} respectively. In this case, the four-dimensional
Lagrangian is reduced to%
\begin{eqnarray}
\mathcal{L}_{\text{gWZW}}^{\text{UR}} &=&\kappa \left[ \epsilon _{IJKL}%
\mathcal{R}^{IJ}\mathcal{R}^{KL}\phi +4\epsilon _{IJKL}\mathcal{R}^{IJ}%
\mathcal{R}^{K}\phi ^{L}+i\mathcal{R}^{IJ}\mathcal{R}_{IJ}\varphi \right. 
\notag \\
&&\left. -8\left( \bar{\chi}\mathcal{R}^{IJ}\Gamma _{IJ}\text{D}_{\omega
}\psi +\text{D}_{\omega }\bar{\psi}\mathcal{R}^{IJ}\Gamma _{IJ}\chi -\frac{1%
}{2}\text{D}_{\omega }\bar{\chi}\mathcal{R}^{IJ}\Gamma _{IJ}\text{D}_{\omega
}\chi \right) \right]\,.
\end{eqnarray}%
As before, we shall decompose the five-dimensional spatial index, following
the structure and change of notation of eqs. \eqref{RIJ} and \eqref{Ti}. In
these terms, the four-dimensional ultra-relativistic gWZW Lagrangian is
given by%
\begin{eqnarray}
\mathcal{L}_{\text{gWZW}}^{\text{UR}} &=&\kappa \left[ 4\epsilon _{ijk}%
\mathcal{R}^{ij}T^{k}\phi +8\epsilon _{ijk}T^{i}\mathcal{R}^{j}\phi
^{k}+4\epsilon _{ijk}\mathcal{R}^{ij}\mathcal{R}^{k}\rho -4\epsilon _{ijk}%
\mathcal{R}^{ij}T\phi ^{k}+i\mathcal{R}^{ij}\mathcal{R}_{ij}\varphi
+2iT^{i}T_{i}\varphi \right.  \notag \\
&&-8\left( \bar{\chi}\mathcal{R}^{ij}\Gamma _{ij}\text{D}_{\omega }\psi +2%
\bar{\chi}T^{i}\Gamma _{i}\Gamma \text{D}_{\omega }\psi +\text{D}_{\omega }%
\bar{\psi}\mathcal{R}^{ij}\Gamma _{ij}\chi +2\text{D}_{\omega }\bar{\psi}%
T^{i}\Gamma _{i}\Gamma \chi \right.  \notag \\
&&\left. \left. -\frac{1}{2}\text{D}_{\omega }\bar{\chi}\mathcal{R}%
^{ij}\Gamma _{ij}\text{D}_{\omega }\chi -\text{D}_{\omega }\bar{\chi}%
T^{i}\Gamma _{i}\Gamma \text{D}_{\omega }\chi \right) \right]\,,
\end{eqnarray}%
with the convention $\epsilon _{1234}=1$ and $\phi ^{4}=\rho $.

\section{Concluding remarks}\label{ccl}

In this article, we have obtained a four-dimensional theory for
supergravity. The construction of the action makes use of the
five-dimensional $\mathcal{N}=1$ Poincar\'{e} supergroup as gauge group, a
one form gauge connection evaluated on it, and a transformed gauge
connection in which the gauge parameter is evaluated in the coset space
resulting between the five-dimensional Poincar\'{e} superalgebra and its
Lorentz algebra. The existence of such transformed connection is enough to
formulate five-dimensional standard supergravity as a gauge invariant theory
of the Poincar\'{e} supergroup by means of the SW-GN formalism. This is
carried out by interpreting the new connection as the fundamental field of a
supergravity theory (instead of an equivalent gauge connection associated to
the original one by means of a symmetry transformation), which in this case
means to consider the transformed one-form $V^{A}$ associated to the
translation generators as f\"{u}nfbein. Furthermore, we have also introduced
a transgression field theory that leads to a gWZW model. As it happens in
the SW-GN formalim, the field content of such model is given by the original
one-form gauge connection in addition to the parameters zero-forms. The
resulting action principle is fully gauge-invariant under the Poincar\'{e}
supergroup and corresponds to a supersymmetric extension of the
even-dimensional topological gravity introduced by A. H. Chamseddine in ref. 
\cite{Chamseddine:1990gk}. 

By considering the above-mentioned gWZW
supergravity theory as a starting point, we have studied two specific
regimes, namely, the non- and ultra-relativistic limits. For the first one,
we have found a $\mathcal{N}=1$ supersymmetric extension of the Galilei
algebra in five dimensions. Moreover, we derived a non-linear connection
that allows the formulation of five-dimensional gauge invariant theories by
means of the SW-GN formalism, and also obtained the corresponding
four-dimensional gWZW model. We have found that, in this regime, the
resulting gWZW model is not supersymmetric. Thus, we have obtained a
non-relativistic gravity theory in four dimensions, which is invariant under
the bosonic sector of the mentioned non-relativistic algebra. On the other
hand, for the second case, we have found a five-dimensional supersymmetric
extension of the Carroll algebra, obtained a non-linear realization of it,
and finally derived and a four-dimensional\ gWZW action principle that, in
contrast with the non-relativistic case, preserves supergravity.

It would be interesting to consider the semigroup expansion method \cite{Izaurieta:2006zz} in order to derive a non-relativistic five-dimensional supergravity action. As it was shown in \cite{Concha:2023bly}, some semigroups are useful to derive non-relativistic algebras with a non-degenerate invariant tensor. It would be worth exploring if such particularity can be extended in presence of supersymmetry. In particular, following the examples obtained in three spacetime dimensions \cite{Concha:2020tqx,Concha:2020eam,Concha:2021jos,Concha:2021llq}, one could explore the construction of a trully supersymmetric gravity action in five spacetime dimensions along its dimensional reduction. 

It would also be interesting to extend the study and the analysis done in this work to the case of a non-vanishing cosmological constant in the starting supergravity theory in five-dimensions \cite{Izaurieta:2006wv}. Another aspect that deserves to be studied is the generalization of our results to $\mathcal{N}$-extended supergravities, with and without cosmological constant. Finally, it would be worth extending the analysis done along this work to the case of hypergravity, both in three and five spacetime dimensions \cite{Chen:2013oxa,Zinoviev:2014sza,Henneaux:2015ywa,Fuentealba:2015jma,Fuentealba:2015wza,Fuentealba:2019bgb}.

\section*{Acknowledgements}

The authors would like to thank P. Salgado for enlightening discussions.
P.C. acknowledges financial support from the National Agency for Research
and Development (ANID) through Fondecyt grants No. 1211077 and 11220328.
F.I. acknowledges financial support from ANID through Fondecyt grant
1211219. E.R. acknowledges financial support from ANID through SIA grant No.
SA77210097 and Fondecyt grant No. 11220486 and 1231133. P.C. and E.R. would like to
thank to the Direcci\'{o}n de Investigaci\'{o}n and Vicerector\'{\i}a de
Investigaci\'{o}n of the Universidad Cat\'{o}lica de la Sant\'{\i}sima
Concepci\'{o}n, Chile, for their constant support. S.S. acknowledges
financial support from Universidad de Tarapac\'{a}, Chile.

\bibliographystyle{utphys.bst}
\bibliography{Draft-4d}

\end{document}